\documentclass[twocolumn,showpacs,pra,amsmath,amssymb]{revtex4}

\usepackage{graphicx}
\usepackage{dcolumn}
\usepackage{bm}

\begin{document}
\title{Collective motions of a quantum gas confined in a harmonic trap}
\author{Dae-Yup Song}%
 \affiliation{Department of Physics, Sunchon National University,
  Suncheon 540-742, Korea}
\date{\today}
\begin{abstract}
Single-component quantum gas confined in a harmonic potential, but
otherwise isolated, is considered. From the invariance of the system
of the gas under a displacement-type transformation, it is shown
that the center of mass oscillates along a classical trajectory of a
harmonic oscillator. It is also shown that this harmonic motion of
the center has, in fact, been implied by Kohn's theorem. If there is
no interaction between the atoms of the gas, the system in a
time-independent isotropic potential of frequency $\nu_c$ is
invariant under a squeeze-type unitary transformation, which gives
collective {\it radial} breathing motion of frequency $2\nu_c$ to
the gas. The amplitudes of the oscillating and breathing motions
from the {\it exact} invariances could be arbitrarily large. For a
Fermi system, appearance of $2\nu_c$ mode of the large breathing
motion indicates that there is no interaction between the atoms,
except for a possible long-range interaction through the
inverse-square-type potential.
\end{abstract}
 \maketitle

Experiments at ultracold temperature have stimulated theoretical
efforts to explore the properties of quantum many-body systems in
harmonic potentials. As the interaction between the atoms of the
quantum gas is controllable \cite{Fresonance}, the density profile
of $N$-body noninteracting fermions has been investigated from the
eigenstates of a harmonic oscillator \cite{Mueller}. For a
one-dimensional harmonic oscillator, it is known that the
time-dependent Schr\"{o}dinger equation is invariant under the
displacement-type transformation \cite{displacement}, and
squeeze-type transformation (up to a rescaling of time)
\cite{squeeze}. The invariance under the displacement-type
transformation (DTT) gives wave functions whose probability
distributions move along classical solutions. The invariance under
the squeeze-type transformation (STT) gives breathing motions to the
distributions. For a time-independent potential of frequency
$\nu_c$, it is numerically shown that the breathing motion has
frequency $2\nu_c$ \cite{PMCLZ}.

Harmonic motions of the centers of masses of the quantum gases have
been clearly noticed in experiments and used to accurately measure
the potentials (see, e.g., \cite{GJ}). Theoretically, dipole mode in
the motion of the center has been found through various
approximations, mainly based on the Gross-Pitaevskii (GP) mean-field
formalism \cite{PMCLZ,Stringari}. In addition, for the
time-independent potential, Kohn's theorem says the existence of
{\it exact} excited states with excitation energies $2\pi
l\hbar\nu_c$  $(l=1,2,\cdots)$  and further "harmonic-potential
theorem" has been established \cite{Kohn}. To my knowledge, however,
the harmonic motion of the center has rather been understood from
the insight that "collisions between atoms cannot alter
center-of-mass momentum" \cite{GJ} in the context of ultracold
temperature physics. The implication of the breathing motion of the
harmonic oscillator to the Bose-Einstein condensates has also long
been studied in literature. In particular, it has been suggested
that the GP equation of a time-dependent harmonic trap may be
transformed to the equation of a time-independent potential through
the unitary transformation of a harmonic oscillator \cite{KSS}, and
a scheme to understand the $2\nu_c$ mode is proposed
\cite{Pitaevskii}. Recently, a $2\nu_c$ mode has been found in an
experiment \cite{MSKE}.

In this paper, single-component quantum gas confined in a
time-dependent harmonic potential, but otherwise isolated, is
considered, with an assumption that position-dependent part of the
interaction between the atoms of the gas is written in terms of the
differences of positions of two atoms. From the {\it exact}
invariance of the gas system under a DTT, it is shown that the
center of mass oscillates along a classical trajectory of a harmonic
oscillator as it has been known for the time-independent potential
through a different formalism \cite{Kohn}. It is also shown that
this harmonic motion of the center has, in fact, been implied by
Kohn's theorem. If the interaction between the atoms is absent, the
system in a time-independent isotropic $\nu_c$ potential is
invariant under a STT up to a rescaling of time, which predicts the
collective {\it radial} breathing motion of frequency $2\nu_c$ of
the gas. The amplitudes of the oscillating and breathing modes from
the {\it exact} invariances could be arbitrarily large, as they are
determined purely by the classical solutions. In this respect, the
{\it exact} modes may be different from other modes found in the
linear approximations \cite{Stringari}. For a Fermi system,
appearance of $2\nu_c$  mode of the large breathing motion signals
that the gas is in the region of no interaction between the atoms,
except for a possible long-range interaction through the
inverse-square-type potential. The collective harmonic motions of
the fermions will show that the complete set of the numbered
coherent states \cite{Nieto,harmonic-Song,CS-Song} of a harmonic
oscillator can be used in stacking the fermions, with equal
validity, as the eigenstates are used.  For simplicity, the
nonlinear Schr\"{o}dinger (generalized GP) equation will be
considered first, and the formalism will be extended to the
many-body systems.

A $D$-dimensional nonlinear Schr\"{o}dinger equation (NLSE) with a
time-dependent harmonic potential is given by
\begin{equation}
O(t,w(t))\Psi(\vec{r},t)+g|\Psi(\vec{r},t)|^{2n}\Psi(\vec{r},t)=0,
\end{equation}
with
\begin{equation}
O(t,w(t))=-i\hbar{\partial \over \partial t}
    -{\hbar^2\vec{\nabla}^2 \over 2m}
    +{m \over 2}\sum_{i=1}^Dw_i^2(t)x_i^2,
\end{equation}
where $\vec{\nabla}=({\partial \over
\partial x_1},{\partial \over
\partial x_2},\cdots{\partial \over
\partial x_D})$, $m$ is the mass of an atom, and $g$,$n$ are constants.
When $f_i$ satisfies the Hill's equation, the classical equation of
motion of a harmonic oscillator,
\begin{equation}
\ddot{f}_i+w_i^2(t)f_i=0,
\end{equation}
with the overdot denoting differentiation with respect to $t$, the
displacement-type unitary operator for the $D$-dimensional harmonic
oscillator can be given as
\begin{equation}
U_f(\vec{r},t)= \prod_{i=1}^D \left( e^{\left[{i\over
\hbar}\left(\delta_i+m\dot{f}_i x_i\right)\right]}
     \exp\left[-f_i {\partial\over \partial x_i} \right]
\right),
\end{equation}
where $\delta_i$ is defined through the relation
\begin{equation}
\dot{\delta}_i={1\over 2}m\left(w_i^2(t)f_i^2-\dot{f}_i^2 \right).
\end{equation}
Making use of the operator, one can find that
\begin{eqnarray}
\tilde{\Psi}(\vec{r},t)&\equiv&U_f(\vec{r},t)\Psi(\vec{r},t),\nonumber\\
 &=&\left(\prod_{i=1}^D
 \exp\left[{i\over\hbar}(\delta_i+m\dot{f}_ix_i)\right]\right)
 \Psi(\vec{r}-\vec{f},t),~~
\end{eqnarray}
with $\vec{f}=(f_1,f_2,\cdots,f_D).$ If the unitary operator
$U_f(\vec{r},t)$ is applied on (1), from the relation
\begin{equation}
U_f(\vec{r},t) O(t,w(t)) U_f^\dagger(\vec{r},t)=O(t,w(t))
\end{equation}
and the fact that
$U_f(\vec{r},t)\left[|\Psi(\vec{r},t)|^2\Psi(\vec{r},t)\right]=
 |\tilde{\Psi}(\vec{r},t)|^2\tilde{\Psi}(\vec{r},t)$, one can find that the transformed
equation is the same one with the replacement of $\Psi(\vec{r},t)$
by $\tilde{\Psi}(\vec{r},t)$. If $\Psi(\vec{r},t)$ is a solution of
(1), $\tilde{\Psi}(\vec{r},t)$ is also a solution of the same
equation, and the number density $|\tilde{\Psi}(\vec{r},t)|^2$ of
the transformed solution has harmonic motions with respect to the
original one, without changing the shape. If $w_i(t)$ is a constant
$w_{ci}$ and the transformation is applied on a stationary solution
of (1), then the number density of the transformed solution
oscillates sinusoidally along the $i$-th direction. For $D=1$, the
invariance has been known through a different formalism \cite{cmn}.

The invariance under the DTT can be easily extended for a $N$-body
system, if the interaction
$V(\vec{r}(1),\vec{r}(2),\cdots,\vec{r}(N))$ between the atoms
satisfies
\begin{eqnarray}
&&V({\vec{r}}(1)+\vec{c},\vec{r}(2)+\vec{c},\cdots,\vec{r}(N)+\vec{c})
\nonumber\\
&&=V(\vec{r}(1),\vec{r}(2),\cdots,\vec{r}(N)),
\end{eqnarray}
where $\vec{r}(j)$ denotes the position of the $j$-th atom and
$\vec{c}$ is a constant vector. By defining $U_N$ as
\begin{equation}
U_N=\prod_{j=1}^NU_f(\vec{r}(j),t)
\end{equation}
one may find the relation
\begin{equation}
U_N O_N(t,w(t)) U_N^\dagger=O_N(t,w(t)),
\end{equation}
where
\begin{eqnarray}
&&O_N(t,w(t))\equiv-i\hbar{\partial \over \partial t}+H_N \nonumber\\
&&=-i\hbar{\partial \over \partial
t}+V(\vec{r}(1),\vec{r}(2),\cdots,\vec{r}(N))
\nonumber\\
&&~+\sum_{j=1}^N \left[
    -{\hbar^2\vec{\nabla}^2(j) \over 2m}
    +{m \over 2}\sum_{i=1}^Dw_i^2(t)x_i^2(j)\right].
\end{eqnarray}
For a given time, the operations of $U_N$ and $U_f(\vec{r},t)$
amount to moving the space coordinate along the vector $\vec{f}$ to
find new solutions, up to the multiplicative phase factors. The
probability distribution of the transformed wave function of the
$N$-body system will thus have the harmonic motion with respect to
the distribution of the original wave function, as the number
density of the transformed solution of the NLSE does.~(10) is valid
for both of the Fermi-Dirac and Bose-Einstein statistics, and is an
{\it exact} explanation for the longstanding observation that the
center of mass of the single-component gas confined in a harmonic
potential moves harmonically, while the harmonic motion of the
center has been known for the time-independent potential
\cite{Kohn}.

For the case of constant frequency $w_i(t)=w_{ci}$ $(i=1,2, \cdots,
D)$, one may find that the harmonic motions of the centers of masses
have, in fact, been implied by Kohn's theorem \cite{Kohn}. For this,
by letting
\begin{equation}
f_i=\sqrt{2\hbar \over mw_{ci}}z_{0i}\cos(w_{ci} t+\varphi_i)
\end{equation}
with real constants $z_{0i},~\varphi_i$, one can find the relations
\begin{eqnarray}
&&\exp\left[{i\over \hbar}\delta_i+{i\over
\hbar}m\dot{f}_ix_i(j)\right]\exp\left[-f_i{\partial \over \partial
x_i(j)}\right]
\nonumber\\
&&~=e^{i\phi_i}\exp\left[{i\over \hbar}m\dot{f}_ix_i(j)
-f_i{\partial \over
\partial x_i(j)}\right]\\
&&~=e^{i\phi_i}\exp[z_i(t)a^\dagger_i(j)-z^*_i(t)a_i(j)],
\end{eqnarray}
where
\begin{eqnarray}
&&z_i(t)=z_{0i}e^{-i(w_{ci}t+\varphi_i)}, \\
&&a_i(j)=\frac{1}{\sqrt{2}}\left(\sqrt{mw_{ci} \over\hbar}x_i(j)
+\sqrt{\hbar\over mw_{ci}}{\partial\over \partial x_i(j)}\right),~~~
\end{eqnarray}
and $z^*_i(t)$ denotes complex conjugate of $z_i(t)$. In (15,16),
$\phi_i$ is a real constant coming from that $\delta_i$ is defined
up to a constant, and from now on we will set $\phi_i=0$ $(i=1,2,
\cdots, D)$. One can thus find that $U_N$ is written as
\begin{eqnarray}
U_N&=&\prod_{i=1}^D\exp\left(z_i(t)A^\dagger_i-z_i^*(t)A_i\right)\\
&=&\prod_{i=1}^D e^{-Nz_{0i}^2/2}\exp\left(z_i(t)A^\dagger_i\right)
\exp\left(-z_i^*(t)A_i\right),~~~~
\end{eqnarray}
where
\begin{equation}
A_i=\sum_{j=1}^Na_i(j).
\end{equation}
Indeed, in Ref.~\cite{FR}, it has been proven that
\[  [H_N, A^\dagger_i]=\hbar w_{ci}A^\dagger_i.\]
If $|G>$ is the many-body ground state of $H_N$ with energy
eigenvalue $E_G$, a wave function of the system is given as
\begin{eqnarray}
&&e^{-\frac{i}{\hbar}E_Gt}U_N|G>\nonumber\\
&&=e^{-\frac{i}{\hbar}E_Gt}\prod_{i=1}^De^{-Nz_{0i}^2/2}
\left(\sum_{l=0}^\infty
\frac{z_i^l}{l!}\left(A_i^\dagger\right)^l|G>\right),~~~~~
\end{eqnarray}
while, as suggested  by Kohn's theorem \cite{Kohn},
$\left(A_i^\dagger\right)^l|G>$ is an eigenstate of the $H_N$ with
the eigenvalue $E_G+l\hbar w_{ci}$ \cite{FR}.

From now on, {\it isotropic} potentials will only be considered, so
that $w_i(t)=w(t)$ for all $i$. For the STT, the unitary operator
\begin{eqnarray}
&&U_s \nonumber\\
&&=\left(\Omega \over mw_c \eta^2(t)\right)^{D\over 4}
~~\exp\left[{i \over 2\hbar}{m\dot{\eta}(t)\over \eta(t)}\vec{r}^2\right] \nonumber\\
&&~~~~~~~~~~~~\times
     \exp\left[-{1\over 2}\left(\ln {mw_c \eta^2(t) \over
         \Omega}\right)\vec{r}\cdot\vec{\nabla}\right]~~
\end{eqnarray}
is introduced, with positive constants $\Omega$ and $w_c$
$(=2\pi\nu_c)$. For the time being, it will be assumed that
$\eta(t)$ is an arbitrary smooth positive function of $t$. If $U_s$
is applied on $\chi(\vec{r},\tau(t))$, the transformed function will
be
\begin{eqnarray}
 \Psi(\vec{r},t)&=&U_s ~\chi(\vec{r},\tau(t))\nonumber\\
 &=& \left(\Omega\over mw_c\eta^2(t)\right)^{D\over 4}
 \exp\left[{i \over 2\hbar}{m\dot{\eta}(t)\over \eta(t)}\vec{r}^2\right]\nonumber\\
 &&\times
 \chi(\sqrt{\Omega\over mw_c}{\vec{r} \over \eta(t)},\tau(t)).~~~~~~
\end{eqnarray}
A NLSE with a time-independent harmonic potential is given
as
\begin{equation}
O_c(\tau(t),w_c)\chi(\vec{r},\tau(t))+g|\chi(\vec{r},\tau(t))|^{2n}\chi(\vec{r},\tau(t))=0,
\end{equation}
where
\begin{equation}
O_c(\tau(t),w_c)=-i\hbar{\partial \over \partial
\tau}-{\hbar^2\vec{\nabla}^2 \over 2m}+{mw_c^2 \over 2}\vec{r}^2.
\end{equation}
From the fact
\begin{eqnarray}
&&U_s\left(|\chi(\vec{r},\tau)|^{2n} \chi(\vec{r},\tau)\right)
\nonumber\\
&&=\left(\Omega\over mw_c\eta^2(t)\right)^{-{nD \over 2}}
|\Psi(\vec{r},t)|^{2n}\Psi(\vec{r},t),
\end{eqnarray}
one may find that the NLSE of (23) can be transformed, through the
STT, to that of (1)
only when
\begin{equation}
{dt\over d\tau}={mw_c\eta^2 \over \Omega}.
\end{equation}
Further, if $\eta(t)$ satisfies
\begin{equation}
m\ddot{\eta}+mw^2(t)\eta-{\Omega^2\over m\eta^3}=0,
\end{equation}
one can find the relation
\begin{equation}
U_sO_c(\tau(t),w_c)U_s^\dagger
 = \left({dt\over d\tau}\right)O(t,w(t))
\end{equation}
which is well-known for the one-dimensional harmonic oscillators
\cite{squeeze,CS-Song}. When $\tau$ satisfies (26), by applying the
operator $U_s$ on (23), one can find that the NLSE is transformed as
\begin{eqnarray}
&&U_s\left[
     O_c(\tau(t),w_c)\chi(\vec{r},\tau(t))+g|\chi(\vec{r},\tau(t))|^{2n}\chi(\vec{r},\tau(t))
     \right]\nonumber\\
&&={mw_c\eta^2(t) \over \Omega}\nonumber\\
&&~~\times\left[
\begin{array}{l}
 O(t,w(t))\Psi(\vec{r},t)\\
  +g\left(\Omega\over mw_c\eta^2(t)\right)^{1-{nD \over 2}}
  |\Psi(\vec{r},t)|^{2n}\Psi(\vec{r},t)
\end{array} \right]=0.~~~~~~
\end{eqnarray}
Through the STT, (23) is thus transformed to (1) when
\begin{equation}
nD=2.
\end{equation}
As far as a NLSE is concerned, this condition is identical to that
given in Ref.~\cite{Pitaevskii}, found in a different context using
a trial function. For $n=1,2$, the squeeze-type relation (29) has
been suggested in Ref. \cite{KSS}.

In the harmonic oscillator, it is known that $\eta(t)$ and a
constant $\Omega$ are written as (see, for example,
Refs.~\cite{squeeze,harmonic-Song})
\begin{equation}
\eta(t)=\sqrt{u^2(t)+v^2(t)},~~\Omega=m[\dot{v}(t)u(t)-
\dot{u}(t)v(t)],
\end{equation}
with two linearly independent real solutions $u(t)$ and $v(t)$ of
\begin{equation}
\ddot{x}_{cl}+w^2(t)x_{cl}=0.
\end{equation}
From now on, we only consider $\tau(t)$ and $\eta(t)$ satisfying
(26,27). When (30) is satisfied, the unitary relations can be used
to find a solution $\Psi(\vec{r},t)$ of (1) from a known solution
$\chi(\vec{r},\tau(t))$ of (23), as
\begin{equation}
\Psi(\vec{r},t)=U_f(\vec{r},t) U_s\chi(\vec{r},\tau(t)).
\end{equation}
If $e^{-i\mu\tau(t)}\chi_0(\vec{r})$ is a solution of (23) with a
constant $\mu$, a solution of (1) is given as
\begin{eqnarray}
 &&\Psi(\vec{r},t)\nonumber\\
 &&=\left(\Omega\over mw_c\eta^2(t)\right)^{D\over 4}
  \left({u(t)-iv(t) \over \eta(t)}\right)^{\mu/\hbar}\nonumber\\
 &&~~\times \left(\prod_{i=1}^D
 \exp\left[{i\over\hbar}(\delta_i+m\dot{f}_ix_i)\right]\right)
\nonumber\\
 &&~~\times\exp\left[{im\dot{\eta}(t) \over 2\hbar\eta(t)}(\vec{r}-\vec{f})^2\right]
 \chi_0(\sqrt{\Omega\over mw_c}{\vec{r}-\vec{f} \over
 \eta(t)}).~~~~~
\end{eqnarray}
For the case of $\vec{f}=0$, one finds that
\begin{equation}
|\Psi(\vec{r},t)|^2=\left(\sqrt{\Omega\over
mw_c}{1\over\eta(t)}\right)^D |\chi_0(\sqrt{\Omega\over
mw_c}{\vec{r} \over \eta(t)})|^2.
\end{equation}
For a given time with $\vec{f}=0$, (35) shows that the number
density of $\Psi(\vec{r},t)$ is given from that of
$e^{-i\mu\tau(t)}\chi_0(r)$ by globally rescaling the space
coordinate along the radial direction, with a multiplication factor
needed to keep the total number of particles.

The squeeze-type unitary relation can be extended for the
interacting $N$-body system, if interaction between the atoms is of
homogeneous degree $-2$, so that
\begin{equation}
V(a \vec{r}(1),a \vec{r}(2),\cdots,a\vec{r}(N))= a^{-2}V(\vec{r}(1)
\vec{r}(2),\cdots,\vec{r}(N))
\end{equation}
with a constant $a$. By defining
\begin{eqnarray}
&&O_{N,c}(\tau(t),w_c)\nonumber\\
&&=-i\hbar{\partial \over\partial\tau}
+V(\vec{r}(1),\vec{r}(2),\cdots,\vec{r}(N))
\nonumber\\
&&~~+\sum_{j=1}^N\left[-{\hbar^2\vec{\nabla}^2(j) \over 2m}+{mw_c^2
\over 2}\vec{r}^2(j)\right],~~~~~~~~~
\end{eqnarray}
\begin{eqnarray}
&U_{N,s}& \nonumber\\
    &=&\left(\Omega \over mw_c \eta^2(t)\right)^{ND\over 4}
    \prod_{j=1}^N
     \exp\left[{i \over 2\hbar}{m\dot{\eta}(t)\over \eta(t)}\vec{r}^2(j)\right]
\nonumber\\
&&\times
     \exp\left[-{1\over 2}\left(\ln {mw_c \eta^2(t) \over
         \Omega}\right)\vec{r}(j)\cdot\vec{\nabla}(j)\right],~~
\end{eqnarray}
one may find the relation
\begin{equation}
U_{N,s}O_{N,c}(\tau(t),w_c)U_{N,s}^\dagger
 = \left({dt\over d\tau}\right)O_N(t,w(t)).
\end{equation}
If $\chi_{N}(\tau(t))$ $[=e^{-iE\tau(t)/\hbar}\chi_{N,0}(\vec{r}(1),
\vec{r}(2),\cdots,\vec{r}(N))]$ satisfies the Schr\"{o}dinger
equation
\begin{equation}
O_{N,c}(\tau(t),w_c)\chi_{N}(\tau(t))=0,
\end{equation}
a solution of the Schr\"{o}dinger equation
\begin{equation}
O_N(t,w(t))\Psi_N(t)=0
\end{equation}
is thus given as
\begin{eqnarray}
 &&\Psi_N(t)\nonumber\\
 &&=\left(\Omega\over mw_c\eta^2(t)\right)^{ND\over 4}
  \left({u(t)-iv(t) \over \eta(t)}\right)^{E/\hbar}\nonumber\\
 &&~~\times \left(\prod_{j=1}^N\exp\left[{im\dot{\eta}(t) \over
 2\hbar\eta(t)}\vec{r}^2(j)\right]\right)\nonumber\\
 &&~~\times
 \chi_{N,0}(\vec{\tilde{r}}(1),\vec{\tilde{r}}(2),\cdots,\vec{\tilde{r}}(N)),
\end{eqnarray}
where $\vec{\tilde{r}}(j)=\sqrt{\Omega\over mw_c}{\vec{r}(j) \over
 \eta(t)}.$
As in the number density of the NLSE, for a given time, the
probability distribution of $\Psi_N(t)$ is found from that of
$\chi_{N}(\tau(t))$ by globally rescaling the space coordinate along
the radial direction with a multiplication factor.

When $w(t)=w_c$ (and (30) is satisfied for the NLSE), the
squeeze-type operators in (21,38) transform (23,40) into themselves
with a replacement of $\tau$ by $t$ (up to a rescaling of the time),
respectively, and thus the unitary relations give invariance. If
$\eta(t)$ is denoted as $\eta_c(t)$ in this case, $\eta_c(t)$ is
written without losing generality as
\begin{eqnarray}
&&\eta_c(t)=\sqrt{\Omega\over mw_c} \nonumber\\
&&~~~~~~~~\times\sqrt{A^2\cos^2
w_c(t-t_0)+A^{-2}\sin^2w_c(t-t_0)},~~~~~~
\end{eqnarray}
with a constant $t_0$ and a non-zero constant $A.$ $\eta_c(t)$ is a
periodic function of time with frequency $2\nu_c$, and thus the
probability distribution of the wave function obtained from a
stationary state of the $N$-body system through the STT has the
breathing motion of frequency $2\nu_c$.

The invariance under the STT is possible only when the atoms
interact each other through the potential satisfying (27). Such
potentials are the inverse-square-type \cite{Gambardella} and the
two-dimensional Dirac delta potentials, including other variants
from the one-dimensional Dirac delta potential. Since Dirac delta
potential cannot be used for the interaction of fermions due to the
exclusion principle, the Fermi system with the collective $2\nu_c$
breathing mode should be interpreted as that of no interaction
between the atoms (except for a possible long-range interaction
through the inverse-square-type potential).
In the Tonks-Girardeau limit of the Tonks gas, interacting bosons
behaves like non-interacting quasifermions, and the mean-field
description has been given by Kolomeisky  {\it et al.} \cite{KNSQ}.
Since noninteracting particle systems in one dimension have the
invariance, the mean-field equation should also be invariant under
the STT, and the equation of Kolomeisky {\it et al.} satisfies the
requirement of (30).

In order to realize the large breathing mode from a stationary state
in a time-independent trap of frequency $\nu_c$, one may modulate
$w(t)$ until $\eta(t)$ determined by (27) has a large oscillating
behavior. In this case a convenient choice for the {\it arbitrary}
constant $\Omega$ is $\Omega=mw_c$, and the limiting case of
$A\rightarrow 0$ or $A\rightarrow \infty$ of (43) shows the
existences of such modulations. The modulation of frequency has
already been widely used in experiments (see, e.g., \cite{MSKE}). If
the amplitude of the center-of-mass oscillation decays in an
experiment of a single-component gas using a time-independent
harmonic potential, it indicates that the gas is not completely
isolated. While a $2\nu_c$ breathing mode is found and a sinusoidal
curve is used to fit the size of the breathing gas in
Ref.~\cite{MSKE}, (35,43) imply that the difference between the
radius of the {\it breathing motion from the invariance} and the
sinusoidal curve will be clear for $A>>1$ (or $A<<1$) around the
time the gases are most compressed.

In summary, isolated single-component quantum gas in a harmonic
potential is considered. The invariance under the DTT shows that the
center of mass of the gas moves along a classical trajectory of a
harmonic oscillator. For a time-independent potential, it is shown
that the harmonic motion of the center has, in fact, been implied by
Kohn's theorem \cite{Kohn,FR}. For a Fermi system, through the
invariance under the STT, it is shown that appearance of the large
radial $2\nu_c$ breathing mode indicates that the atoms do not
interact with each other (except for a possible long-range
interaction). While some NLSEs and noninteracting Fermi gas in any
dimension are invariant under the STT, the collective breathing
motion of the one-dimensional gas can be understood from the
evolution of an eigenstate of the harmonic oscillator into a
coherent state when modulation is applied to the frequency.
Alternatively, in one dimension, the collective breathing and
oscillating motions in a time-independent potential can be
interpreted as the result of using generalized coherent states
\cite{Nieto,harmonic-Song,CS-Song} of a harmonic oscillator in
stacking fermions.

\begin{acknowledgments}
The author thanks referees for informing him on Kohn's theorem.
\end{acknowledgments}

\end{document}